\documentclass[preprint,12pt,5p,twocolumn]{elsarticle}

\usepackage{amssymb}

\usepackage{lineno}

\usepackage[hang,small,bf]{caption}
\usepackage[subrefformat=parens]{subcaption}
\usepackage[hidelinks]{hyperref}
\usepackage{flushend}
\usepackage{balance}
\usepackage{upgreek}

\newcommand{\eqref}[1]{Eq.~\ref{#1}}

\usepackage{xspace}
\newcommand{\cmsq}{$\mathrm{cm^2}$\xspace}
\newcommand{\megaohmsq}{$\mathrm{M \Omega/sq}$\xspace}
\newcommand{\freon}{$\mathrm{C_2H_2F_4}$\xspace}
\newcommand{\sfsix}{$\mathrm{SF_6}$\xspace}
\newcommand{\isobutane}{$\mathrm{iC_4H_{10}}$\xspace}

\newcommand{\muegamma}{$\mu \to e \gamma$\xspace}

\journal{Nuclear Instruments and Methods in Physics Research Section A}

\begin{document}

\begin{frontmatter}



\title{Radiation Hardness Studies of RPC Based on Diamond-Like Carbon Electrodes for MEG II Experiment}


\author[kobe-u]{Masato~Takahashi\corref{cor}}
\ead{m.takahashi@stu.kobe-u.ac.jp}
\author[icepp]{Sei~Ban}
\author[u-tokyo]{Weiyuan~Li}
\author[kobe-u]{Atsuhiko~Ochi\fnref{fn1}}
\author[icepp]{Wataru~Ootani}
\author[u-tokyo]{Atsushi~Oya}
\author[kobe-u]{Hiromu~Suzuki}
\author[u-tokyo]{Kensuke~Yamamoto}
\cortext[cor]{Corresponding author}
\fntext[fn1]{Deceased 29 April 2024}

\affiliation[kobe-u]{organization={Department of Physics, Kobe University},
            addressline={1-1 Rokkodai-cho, Nada-ku}, 
            city={Kobe},
            postcode={657-8501}, 
            state={Hyogo},
            country={Japan}}
\affiliation[icepp]{organization={ICEPP, The University of Tokyo},
            addressline={7-3-1 Hongo, Bunkyo-ku}, 
            postcode={113-0033}, 
            state={Tokyo},
            country={Japan}}
\affiliation[u-tokyo]{organization={Department of Physics, The University of Tokyo},
            addressline={7-3-1 Hongo, Bunkyo-ku}, 
            postcode={113-0033}, 
            state={Tokyo},
            country={Japan}}

\begin{abstract}
A novel type of resistive plate chamber, based on diamond-like carbon (DLC) electrodes is under development for background identification in the MEG~II experiment.
The DLC-RPC is required to have 
a radiation hardness to mass irradiation since it is planned to be placed in a high-intensity and low-momentum muon beam.
In this study, the aging test using a high-intensity X-ray beam was conducted to evaluate the radiation hardness of the DLC-RPC.
The accumulated charge due to X-ray irradiation reached about 54~C/\cmsq, which is approximately half of the one-year irradiation dose expected in the MEG~II experiment.
As a result, the degradation of the gas gain was observed due to fluorine deposition and insulators formed on the DLC electrodes.
In addition, discharges via spacers were also observed repeatedly and interrupted the DLC-RPC operation.
\end{abstract}



\begin{keyword}
Gaseous detectors \sep Resistive Plate Chambers \sep Diamond-Like Carbon \sep MEG~II \sep Aging studies


\end{keyword}

\end{frontmatter}



\section{Introduction}
\label{sec:introduction}

The MEG~II experiment \cite{megii-design} searches for \muegamma decay, one of the charged lepton flavor violating processes.
The dominant background is an accidental coincidence between background $\gamma$-rays and background positrons in the MEG~II experiment.
Identification of the radiative muon decay (RMD; $\mu \to e \nu \bar\nu \gamma$), one of the background $\gamma$-ray sources, can further suppress background events.
A novel type of resistive plate chamber, based on diamond-like carbon (DLC) electrodes is under development and planned to be installed in front of the target for further background suppression.
It detects low-energy positrons of 1 -- 5~MeV from the RMD with a high-energy $\gamma$-ray of $E_\gamma>$ 48 MeV. 
The detector should have an ultra-low mass (0.1~\%~$X_0$), high-rate capability (up to 3 MHz/\cmsq), and radiation hardness (irradiation dose of $\sim$~100~C/\cmsq) since the high-intensity ($7 \times 10^7~\mu/\mathrm{s}$) and low-momentum ($28~\mathrm{MeV}/c$) muon beam pass through there. 
Furthermore, the detection efficiency of 90~\% for minimum ionizing particles, the timing resolution of 1 ns, and the 20 cm diameter detector size are also required to identify the RMD efficiently.
The geometrical acceptance to the RMD positrons is expected to be about 100~\% with a 20~cm in diameter detector.

\begin{figure}[tbp]
    \centering
    \includegraphics[width=0.85\linewidth]{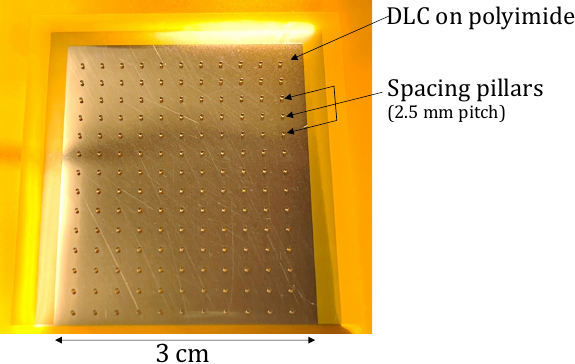}
    \caption{Small prototype electrode of the DLC-RPC. The DLC is sputtered on polyimide and spacing pillars are formed on the DLC.}
    \label{fig:electrode}
\end{figure}

\begin{figure}[tbp]
    \centering
    \includegraphics[width=0.9\linewidth]{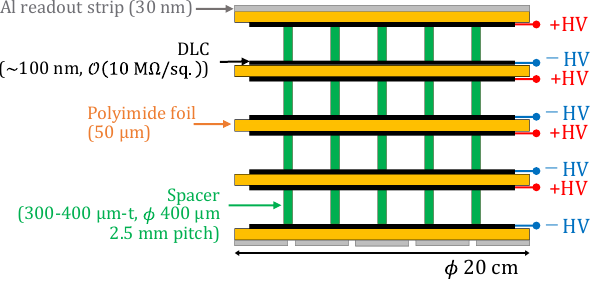}
    \caption{Concept of the full-scale DLC-RPC}
    \label{fig:dlc-rpc}
\end{figure}

%
Performance studies of the DLC-RPC with a 2~cm-size prototype are presented in Ref.~\cite{rpcpaper}.
This prototype demonstrated a low-material design using high-resistivity electrodes (Fig.~\ref{fig:electrode}) made of DLC-coated polyimide films (50~$\upmu$m thick) and 30~nm thick aluminum readout strips.
The use of these materials will allow an RPC of up to four layers within a material budget of 0.1~\%~$X_0$
--- the overall material budget can be suppressed to 0.095~\%~$X_0$ with the design as shown in Fig.~\ref{fig:dlc-rpc}.
For gas gap, on the other hand, 384~$\upmu$m high spacing pillars are formed on the DLC surface by photolithographic processing.
The performances of this prototype detector presented in Ref.~\cite{rpcpaper} are encouraging; 
42~\% single-layer efficiency for MIP particles and 180~ps timing resolution were achieved under a 1~MHz/\cmsq low-momentum muon beam.
When the efficiency of a four-layer RPC is concerned, 42~\% single-layer efficiency is close to the goal of 90~\% four-layer efficiency.
In addition to the above measurements, Ref.~\cite{rpcpaper} finally proposed a further improved detector design, which will give higher rate capability and full scalability of the detector.
The proposed design improvement is underway in parallel with this study to realize a 20~cm size detector.

However, Reference~\cite{rpcpaper} does not address the detector's radiation hardness. Therefore, this study conducted an aging test to evaluate it.



\section{Radiation hardness of the DLC-RPC}
\label{sec:dlc-rpc}

The DLC-RPC is required to have a radiation hardness for a continuous operation for 20~weeks, which corresponds to the duration of the one-year physics run in the MEG~II experiment, under the muon beam.
Aging in gaseous detectors is expected to occur mainly due to the high-energy charge generated by avalanches.
Therefore, the amount of irradiation dose by the muon beam is estimated as the amount of accumulated charge.

Calculated with a beam muon rate of 3~MHz/\cmsq, an average avalanche charge of 3~pC, and an operation period of 20~weeks, the total irradiation dose for MEG~II is expected to be an accumulated charge equivalent to about 100~C/\cmsq.
Here, the average avalanche charge is estimated in the previous test using the muon beam \cite{rpcpaper}.



\section{Setup of aging test using an X-ray beam}
\label{sec:agingtest}

An aging test was conducted using an X-ray generator at KEK in Japan to investigate the radiation hardness of the DLC-RPC.

\subsection{Properties of the X-ray generator}

\begin{figure}[tbp]
    \centering
    \includegraphics[width=0.85\linewidth]{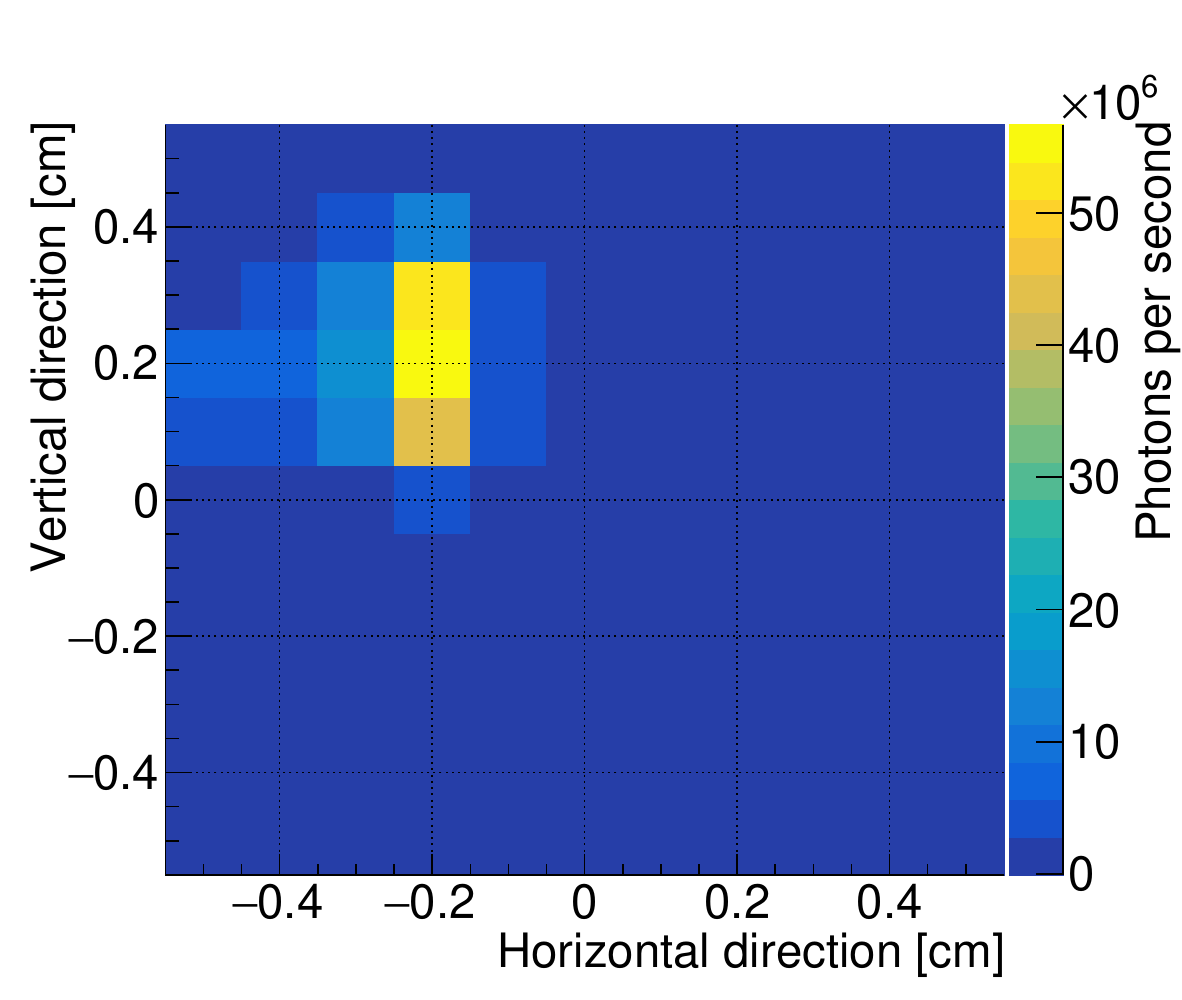}
    \caption{Measured profile of X-ray beams}
    \label{fig:xray_profile}
\end{figure}

This X-ray generator has a copper target and a monochromator.
Therefore, the characteristic X-ray of copper is output, and the energy of the X-ray beam is about 8.04~keV.
The profile of the X-ray beam was measured using an X-ray monitoring device (ionization chamber) as shown in Fig.~\ref{fig:xray_profile}.
X-rays were quite localized to an area of about 1~mm $\times$ 3~mm.
It is locally equivalent to $\sim$~5.7$\times 10^9$~counts/\cmsq of radiation intensity.

\subsection{Experimental setup}

In this test, a small prototype electrode of the DLC-RPC shown in Fig.~\ref{fig:electrode} was used.
As mentioned in Sect.~\ref{sec:introduction}, a polyimide film is coated with DLC, and spacing pillars are formed on DLC at a 2.5 mm pitch by photolithographic technology.
The surface resistivity of the DLC of this prototype is about 40~\megaohmsq.

\begin{figure}[tbp]
    \centering
    \includegraphics[width=0.85\linewidth]{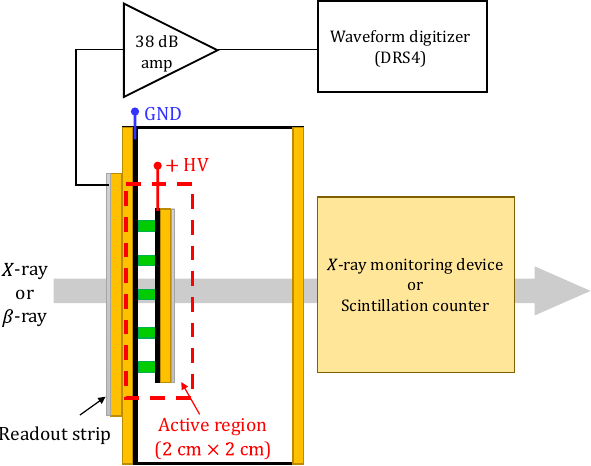}
    \caption{Experimental setup}
    \label{fig:setup}
\end{figure}

Figure~\ref{fig:setup} shows the scheme of the experimental setup. 
The DLC-RPC electrode shown in Fig.~\ref{fig:electrode} was placed on the chamber window, which is a polyimide film with sputtered DLC and works as a cathode electrode of the DLC-RPC.
The surface resistivity of the cathode-side DLC is about 15~\megaohmsq.
The chamber was filled with \freon/\sfsix/\isobutane (94:1:5). 
The signal readout was performed on an aluminum strip, placed on the outer front of the chamber.
The induced signal on the readout strip was amplified by 38~dB amplifiers, and fed into a DRS4 waveform digitizer \cite{drs}.
The X-ray monitoring device was placed behind the chamber to monitor the X-ray beam during X-ray irradiation.
To evaluate changes in the DLC-RPC performance, the response to $\beta$-ray emitted from $^{90}$Sr source was measured before and after X-ray irradiation.
The monitoring device was replaced with a scintillation counter for a reference counter during the $\beta$-ray measurements.

\section{The DLC-RPC status during X-ray irradiation}

\begin{figure}[tbp]
    \centering
    \includegraphics[width=0.9\linewidth]{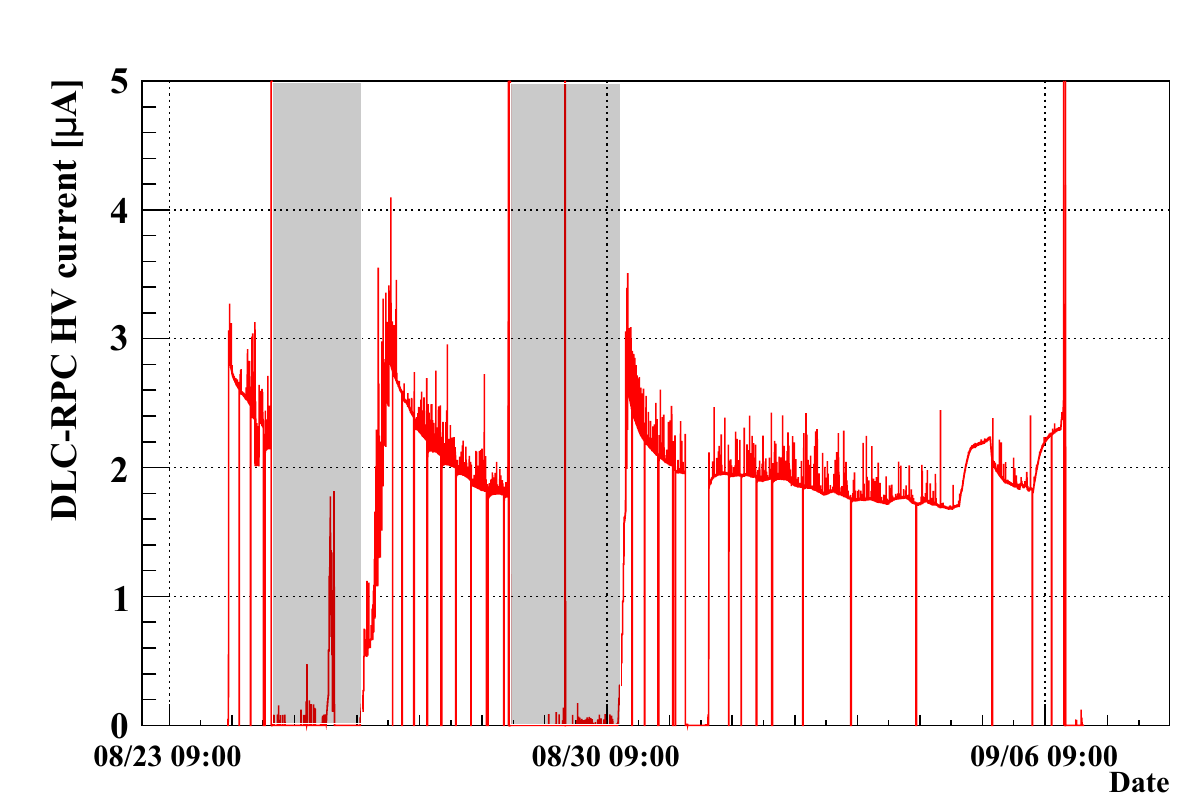}
    \caption{The DLC-RPC HV current. The chamber was opened in shaded periods due to discharge problems.}
    \label{fig:current}
\end{figure}

The DLC-RPC was operated at 2.8~kV of high voltage supply during X-ray irradiation.
The current values of the high-voltage module were continuously recorded to evaluate the accumulated charge.
The history plot of the DLC-RPC HV current is shown in Fig.~\ref{fig:current}.
The operation was interrupted twice because of discharges, which are shown as the shaded region in Fig~\ref{fig:current}.
The chamber was opened and the DLC-RPC electrode was cleaned during these periods for removal of dust that caused discharges via spacers as discussed in Sect.~\ref{sec:discharge}.

Finally, the accumulated charge reached 54~C/\cmsq.
This total charge is approximately half of the irradiation dose expected in the one-year operation in the MEG~II experiment.



\section{Evaluation of aging phenomena}
\label{sec:results}

\subsection{Degradation of the DLC-RPC gas gain}
\begin{figure}[tbp]
    \centering
    \includegraphics[width=0.95\linewidth]{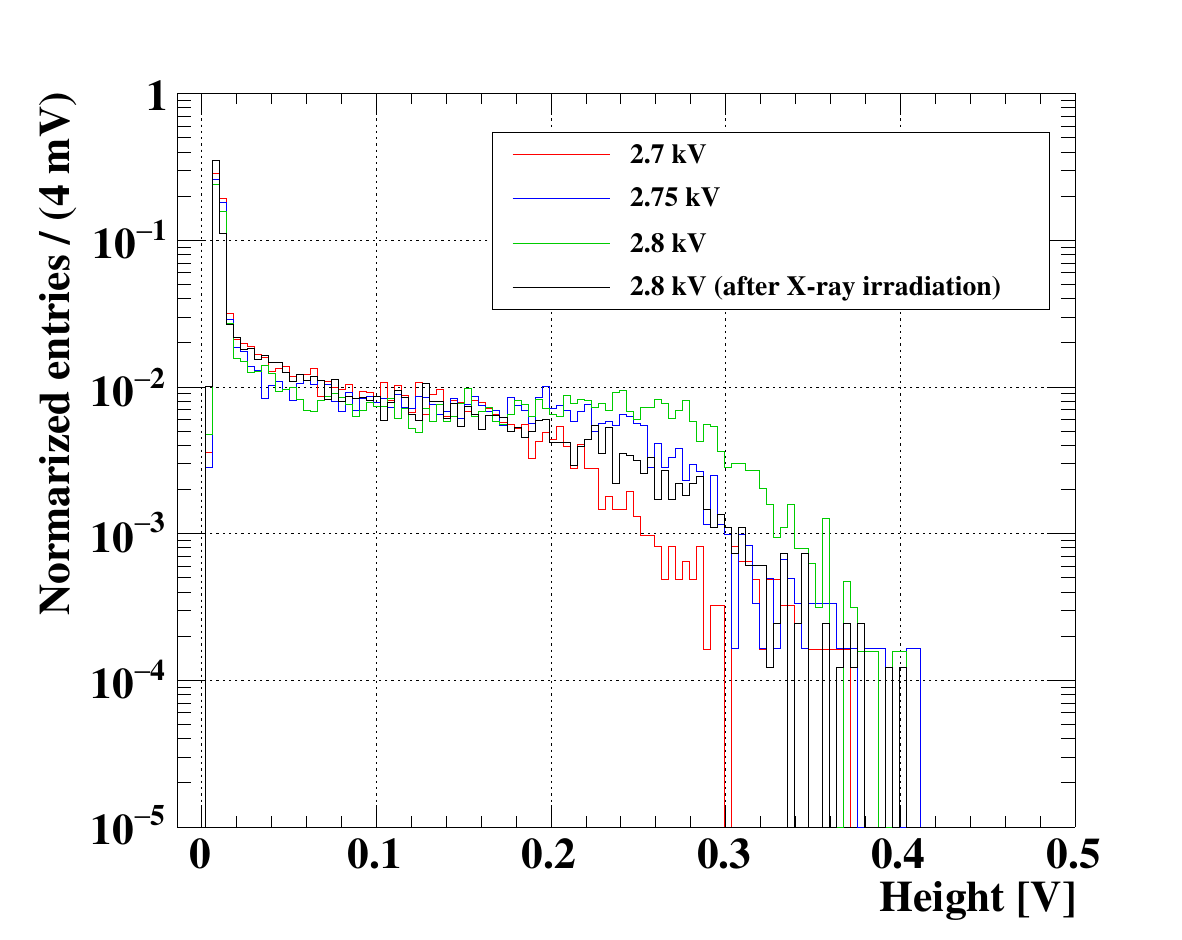}
    \caption{Comparison of the pulse height spectrum for $\beta$-ray}
    \label{fig:spectrum}
\end{figure}

As shown in Fig.~\ref{fig:current}, the DLC-RPC HV current decreased during X-ray irradiation.
This means the decrease in the DLC-RPC gas gain because the HV current was considered proportional to the gas gain.
The X-ray intensity did not change according to the output of the X-ray monitoring device.
Comparison of the pulse height spectra for $\beta$-ray measured before and after the test shows that the effective voltage was reduced by 50 -- 100~V (Fig.~\ref{fig:spectrum}).

\subsection{Deposits on the DLC surface}

\begin{figure}[tbp]
    \centering
    \begin{tabular}{cc}
        \begin{minipage}[b]{0.45\hsize}
            \centering
            \includegraphics[width=0.75\linewidth]{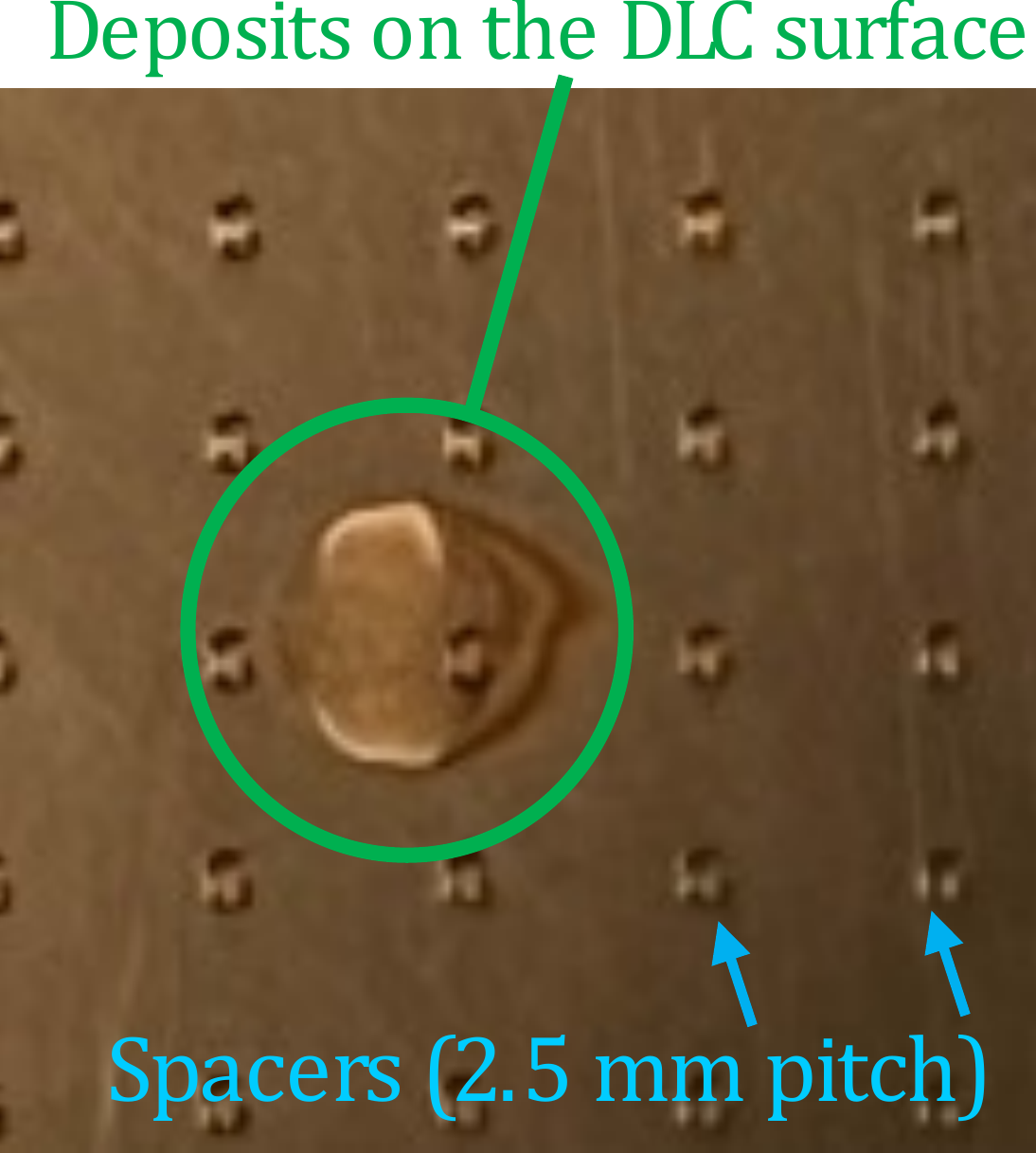}
            \subcaption{}
            \label{fig:discoloration}
        \end{minipage}
        \begin{minipage}[b]{0.45\hsize}
            \centering
            \includegraphics[width=0.85\linewidth]{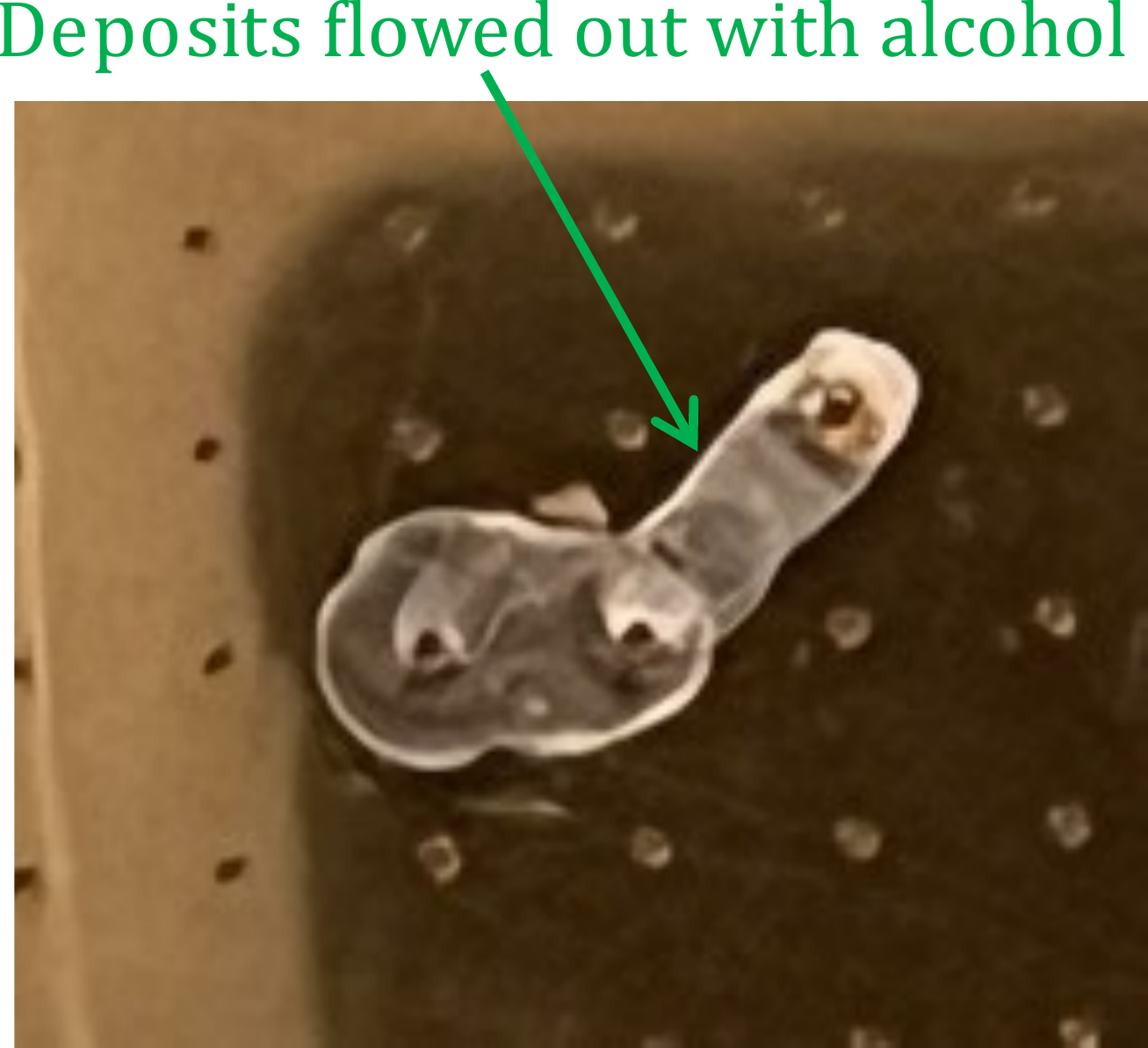}
            \subcaption{}
            \label{fig:flowout}
        \end{minipage}
    \end{tabular}
    \caption{Deposits on the DLC surface. (a) Deposits in the discolored position. (b) Deposits are being removed by alcohol.}
    \label{fig:deposits}
\end{figure}

After X-ray irradiation, deposits on the DLC surface were observed as shown in Fig.~\ref{fig:deposits}.
The resistivity of the DLC was significantly increased in this region ($>$~500~\megaohmsq).
After the removal of the deposits, the resistivity of the DLC and the DLC-RPC gas gain were fully recovered.
We concluded that deposits formed insulators on the DLC surface, causing a decrease in DLC-RPC gas gain.

\subsection{Electrode surface analysis}

\begin{figure}[h]
    \centering
        \begin{minipage}[b]{1.0\hsize}
            \centering
            \includegraphics[width=0.95\linewidth]{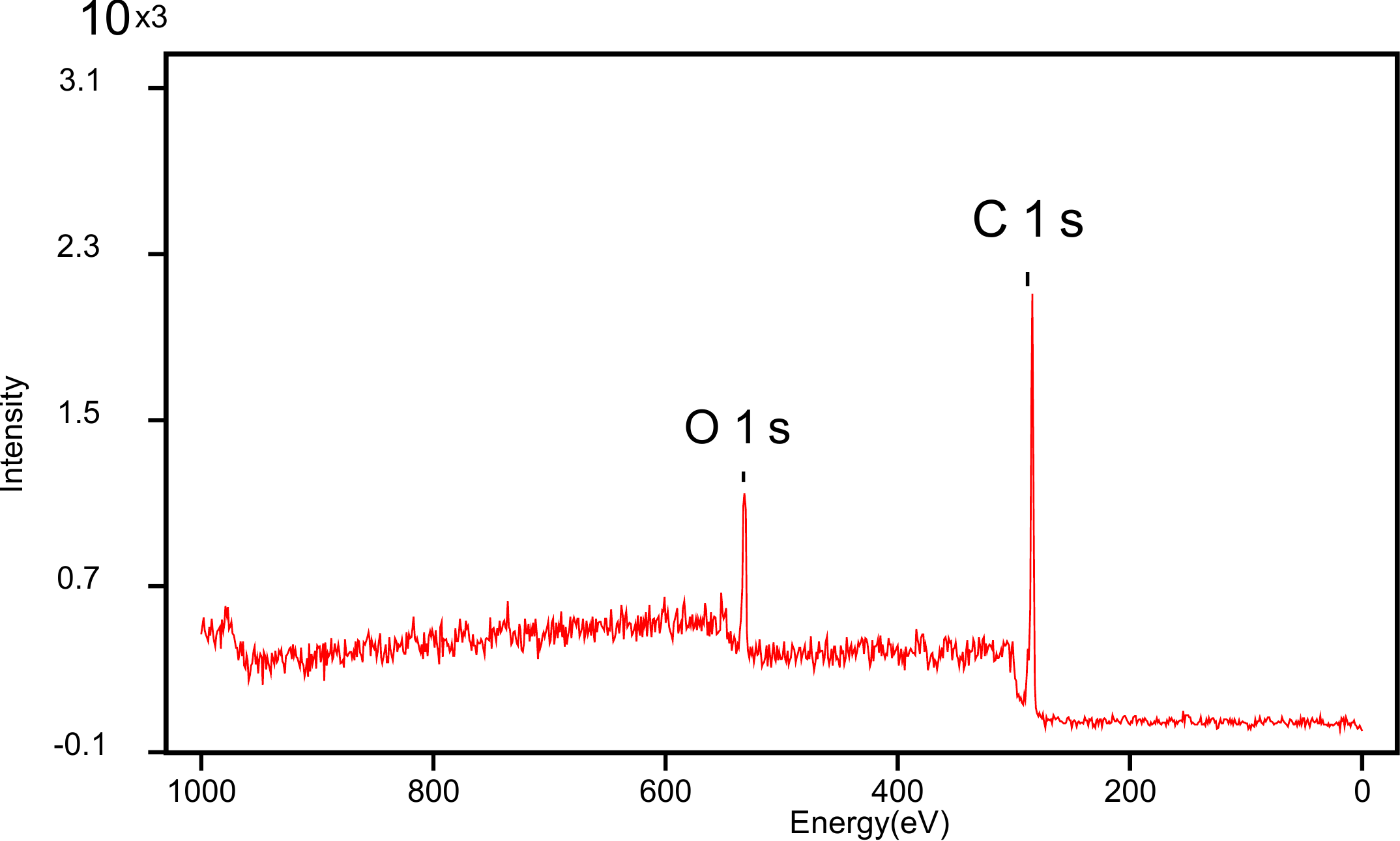}
            \subcaption{Non irradiation area}
            \label{fig:xps_nonirr}
        \end{minipage}\\
        \begin{minipage}[b]{1.0\hsize}
            \centering
            \includegraphics[width=0.95\linewidth]{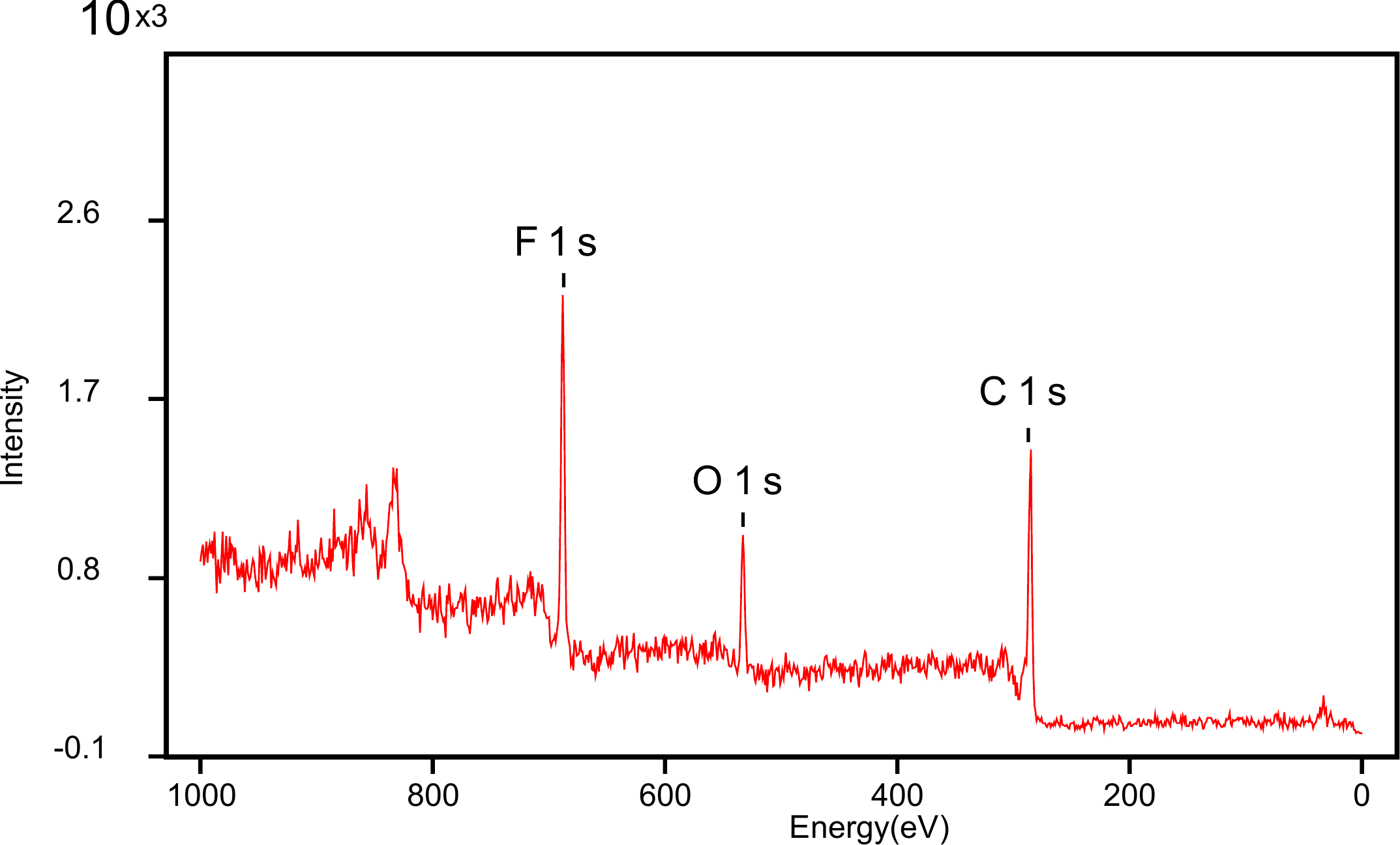}
            \subcaption{Irradiation area}
            \label{fig:xps_irr}
        \end{minipage}
    \caption{Results of the electrode surface analysis}
    \label{fig:xps}
\end{figure}

The chemical composition of the surface deposits was investigated in X-ray photoelectron spectroscopy (PHI X-took, ULVAC-PHI INC. \cite{xps}).
The results are shown in Fig.~\ref{fig:xps} and Tab.~\ref{tab:xps}.
Carbon and oxygen were detected in the non-irradiated area. (Fig.~\ref{fig:xps_nonirr}).
Carbon is a component of the DLC and oxygen is assumed to be detected by oxidation of DLC.
On the other hand, the fluorine component appeared in the X-ray irradiated area (Fig.~\ref{fig:xps_irr}).
This fluorine component is understood to be produced during the irradiation through the following reaction \cite{phan_2017}:
\begin{eqnarray}
    \mathrm{SF_6 + e^-} \to \mathrm{SF_6^{-\ast}}, \quad
    \mathrm{SF_6^{-\ast}} \to \mathrm{SF_5^{-} + F}, \nonumber
\end{eqnarray}
where the electrons are provided by the avalanche clusters.

\begin{table}[tbp]
  \caption{Component ratio}
  \label{tab:xps}
  \centering
  \begin{tabular}{c|cc}
  \hline
  Component & (a) & (b) \\ \hline
  C$_{1\mathrm{s}}$ & 79.66~\% & 63.39~\% \\
  O$_{1\mathrm{s}}$ & 20.34~\% & 11.33~\% \\
  F$_{1\mathrm{s}}$ &  --- & 25.27~\% \\ \hline
  \end{tabular}
\end{table}

\subsection{DLC peeling}

\begin{figure}[tbp]
    \centering
    \includegraphics[width=0.35\linewidth]{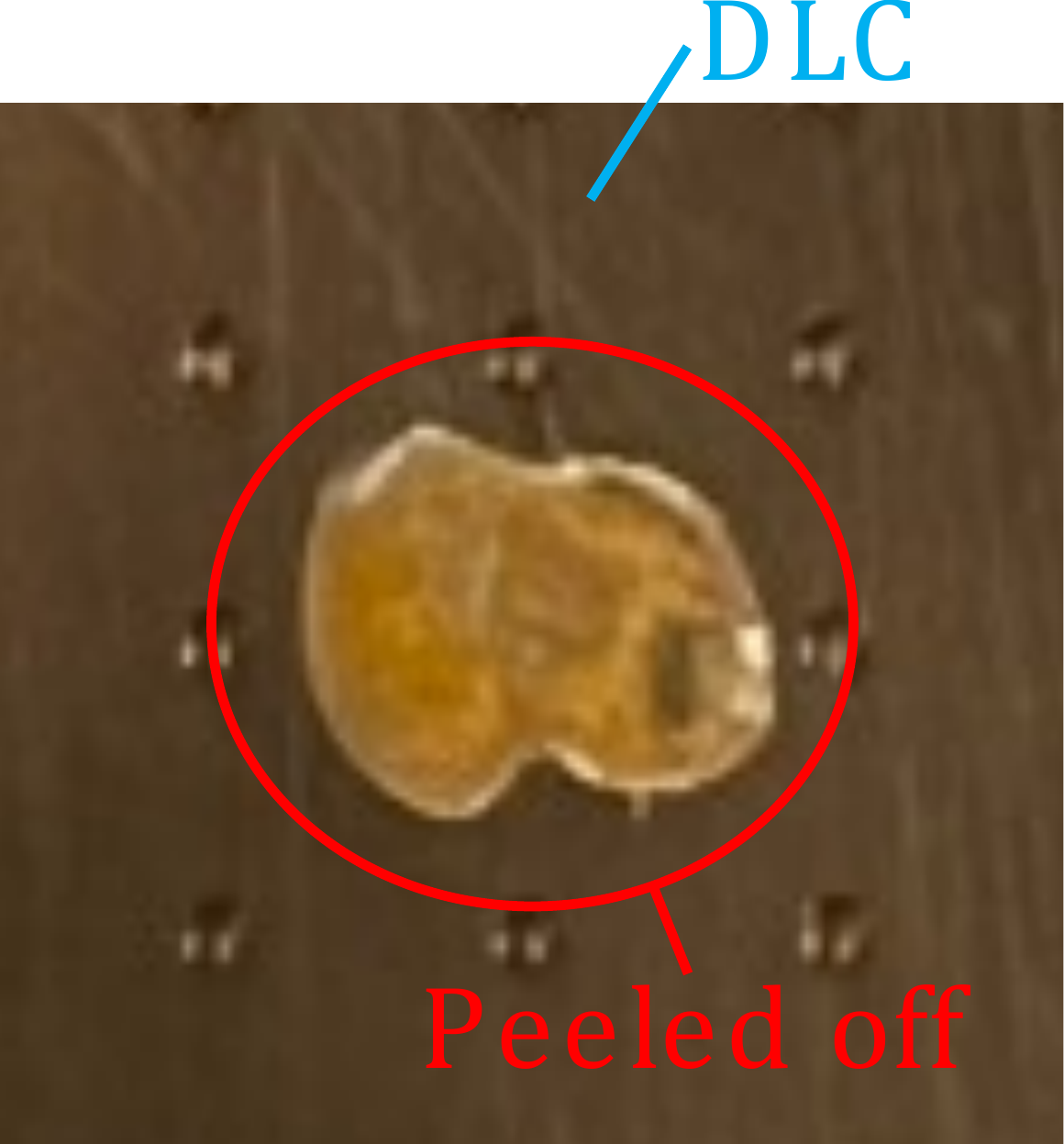}
    \caption{DLC peeling. The orange area in the center has DLC peeling off, exposing the base material polyimide.}
    \label{fig:peeling}
\end{figure}

Although the adhesion between DLC and polyimide is known to be good, the DLC peeling was observed in the X-ray irradiation area as shown in Fig.~\ref{fig:peeling}.
To explain the mechanism, it is supposed that the fluorine separates bonds between the carbons in the DLC. 
However, a clear mechanism is not known, and further investigation is needed.

\subsection{Discharge problems}
\label{sec:discharge}

\begin{figure}[tbp]
    \centering
    \includegraphics[width=0.95\linewidth]{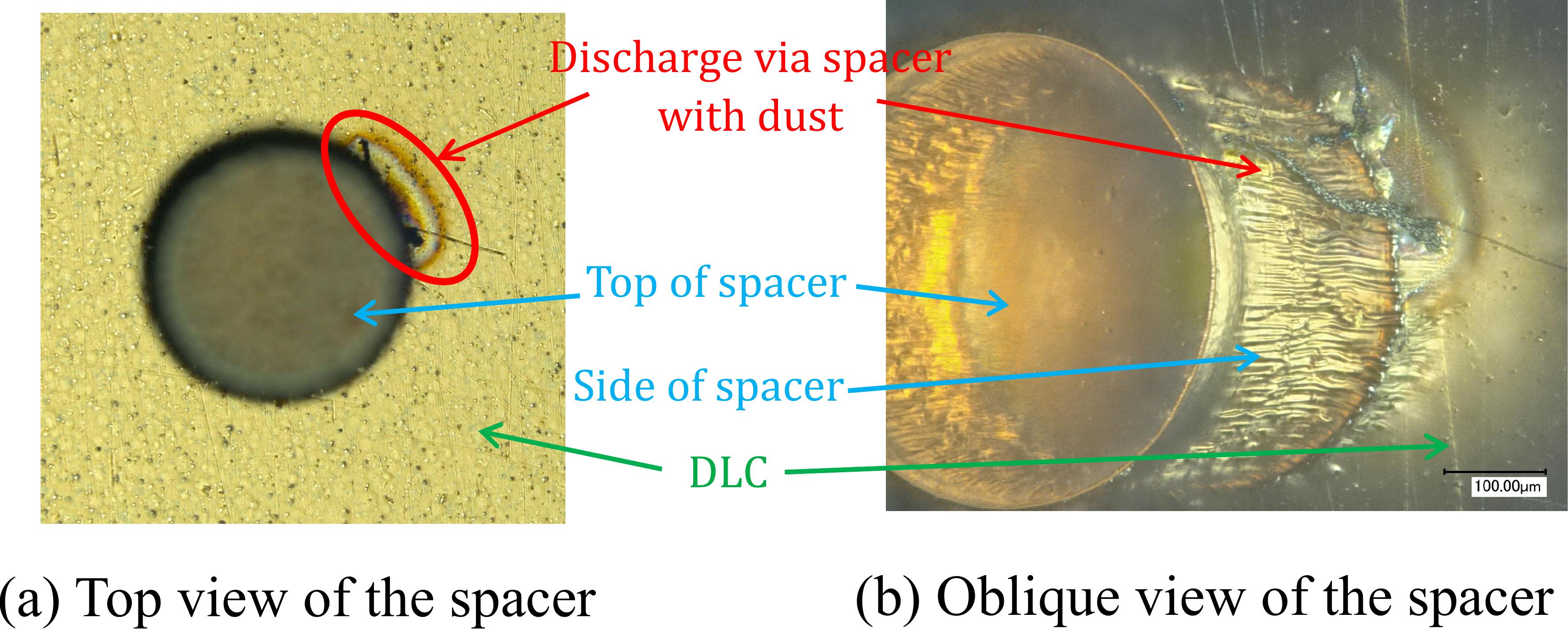}
    \caption{Discharge via a pillar.}
    \label{fig:discharge}
\end{figure}

Twice discharges during the X-ray irradiation (Fig.~\ref{fig:current}) occurred via the spacing pillars.
There is discoloration from discharge at the base of the pillar, and dust on the side of the pillar is presumed to have served as a discharge path as shown in Fig.~\ref{fig:discharge}.
Such discharges occurred repeatedly at the same pillar.
In this test, no further discharge was observed because the relevant pillar was removed after the second discharge.

Discharge via the pillar is a significant problem for the stable operation of the DLC-RPC.
Therefore, it is necessary to establish a mechanism to suppress discharges and a system that enables remote recovery from outside the chamber in the event of a discharge.

\section{Conclusion}
\label{sec:conclusion}

A novel type of resistive plate chamber, based on diamond-like carbon electrodes is under development for background $\gamma$-ray identification in the MEG~II experiment.
Using the small prototype electrode, the radiation hardness of the DLC-RPC was evaluated. 
As a result, the degradation of the DLC-RPC gas gain due to fluorine deposition and interruptions of operation due to discharges were observed.
To solve these problems, future development work should be done to suppress deposit formation, improve quality control, and develop a structure that suppresses discharges.




\section*{Acknowledgement}
\label{sec:acknowledgement}
This work is supported by the KEK Detector R\&D Platform, JSPS KAKENHI Grant Number JP21H04991, JST SPRING Grant Number JPMJSP2148, and Kobe University Research Facility Center for Science and Technology.

\bibliographystyle{elsarticle-num} 
\bibliography{references}





\end{document}